%% file: sample-sigconf-authordraft.tex
\documentclass[sigconf]{acmart} % authordraft
\usepackage{csquotes}
\usepackage{amsfonts,amsmath,amsthm}

\usepackage{float} % in the preamble
\usepackage{multirow} % for \multirow in tables
\usepackage{makecell}
\AtBeginDocument{%
  }

\setcopyright{acmlicensed}
\copyrightyear{2025}
\acmYear{2025}
\acmConference[Preprint]{Preprint}{2025}{Preprint}
\begin{document}

\title{MixLM: High-Throughput and Effective LLM Ranking via Text-Embedding Mix-Interaction}

\author{
Guoyao Li\quad Ran He\textsuperscript{*}\quad Shusen Jing\textsuperscript{*}\quad Kayhan Behdin\textsuperscript{*}\quad Yubo Wang\textsuperscript{*}\quad \\
Sundara Raman Ramachandran\textsuperscript{*}\quad Chanh Nguyen\textsuperscript{*}\quad Jian Sheng\textsuperscript{*}\quad Xiaojing Ma\textsuperscript{*}\quad Chuanrui Zhu\textsuperscript{*}\quad Sriram Vasudevan\quad Muchen Wu\quad Sayan Ghosh\quad Lin Su\quad Qingquan Song\textsuperscript{\dag}\quad Xiaoqing Wang\quad Zhipeng Wang \quad Qing Lan \quad Yanning Chen\quad Jingwei Wu\quad Luke Simon\textsuperscript{\dag}\quad Wenjing Zhang\quad \\
Qi Guo\textsuperscript{\ddag}\quad Fedor Borisyuk\textsuperscript{\ddag}\quad \\
\texttt{\{guoyli, rahe, sjing, kbehdin, yubwang, suramachandran, cnguyen, jsheng, xjma, chuzhu, svasudevan, muwu, sayaghosh, lsu, qsong, xiaoqwang, zhipwang, qlan, yannchen, jingwu, lsimon, wzhang, qguo, fborisyuk\}@linkedin.com} \\
LinkedIn, Mountain View, CA, USA
}

\renewcommand{\shortauthors}{Guoyao Li et al.}

\begin{abstract}
Large language models (LLMs) capture rich semantic signals and achieve strong relevance ranking
performance in modern search and recommendation systems, but their high computational cost
poses challenges under industrial latency and throughput constraints. In particular,
cross-encoder rankers incur prefill-heavy inference due to long contexts combining user,
query, and item text. We propose \textbf{MixLM}, an LLM-based ranking framework that
substantially improves throughput by reducing input context length while preserving the
semantic strength of cross-encoder interaction. MixLM introduces a \textbf{mixed-interaction}
representation that combines text tokens with learned embedding tokens. Item descriptions
are pre-encoded into a small number of embedding tokens and stored in a nearline cache, reducing
online item context from thousands of text tokens to a few embeddings at inference time.
We describe the end-to-end training pipeline and serving optimizations for deploying MixLM in
a production LinkedIn search system. Under a fixed latency budget and comparable relevance
metrics, MixLM improves throughput by 10.0$\times$ over strong baselines and by 75.9$\times$
over full-text LLM rerankers. These efficiency gains enabled full-traffic deployment of
LLM-powered search, resulting in a 0.47\% increase in Daily Active Users (DAU) in online A/B
experiments.
\end{abstract}

% <concept_id>10010147.10010257.10010293.10010294</concept_id>
% <concept_desc>Computing methodologies~Neural networks</concept_desc>
% <concept_id>10002951.10003317.10003347.10003350</concept_id>
% <concept_desc>Information systems~Recommender systems</concept_desc>
% <concept_id>10002951.10003317.10003338.10003343</concept_id>
% <concept_desc>Information systems~Learning to rank</concept_desc>

% \ccsdesc[500]{Computing methodologies~Neural networks}
% \ccsdesc[500]{Information systems~Recommender systems}
% \ccsdesc[500]{Information systems~Learning to rank}

\begin{CCSXML}
<ccs2012>
  <concept>
    <concept_id>10010147.10010178.10010224.10010226</concept_id>
    <concept_desc>Computing methodologies~Information retrieval</concept_desc>
    <concept_significance>500</concept_significance>
  </concept>
  <concept>
    <concept_id>10010147.10010178.10010219.10010223</concept_id>
    <concept_desc>Computing methodologies~Natural language processing</concept_desc>
    <concept_significance>500</concept_significance>
  </concept>
</ccs2012>
\end{CCSXML}

\ccsdesc[500]{Computing methodologies~Information retrieval}
\ccsdesc[500]{Computing methodologies~Natural language processing}
% \keywords{Natural language processing, Large language models, recommender systems}
\keywords{
Semantic Search,
Large Language Models,
Learning to Rank
}
\maketitle
\renewcommand{\thefootnote}{\fnsymbol{footnote}}
\footnotetext[1]{Equal key contributors.}
\footnotetext[2]{Work done while at LinkedIn.}
\footnotetext[3]{Correspondence authors.}

\input{introduction}

\section{Semantic Search}

Semantic search retrieves and ranks items based on semantic similarity rather than lexical
overlap. A standard pipeline consists of dense representation learning, large-scale vector
retrieval (e.g., GPU-accelerated exhaustive search), high-capacity reranking to model
query--item interactions, and downstream business logic such as personalization and filtering.

We focus on the \emph{ranking} component and adopt a pointwise formulation in which a model
consumes the query, candidate item, and optional user context to predict relevance or user
engagement. Relevance labels are defined by product policy, and models are trained to optimize
NDCG@10 using policy supervision.

Given a query $q$ and candidate item $i$, a cross-encoder ranker constructs a single textual
prompt
\begin{equation}\label{eq:promt}
    \text{prompt}(q,i) = \text{system prefix}, q, i ,
\end{equation}
where the system prefix includes instructions guiding the LLM to assess semantic match.

Since relevance is treated as a binary classification task%
\footnote{We do not consider setwise or listwise ranking in this work.},
the prompt is processed by an LLM with a classification head that estimates
\begin{equation}
    p_{\text{yes}}(q,i) = \mathbb{P}(i~\text{is relevant to}~q) \in [0,1],
\end{equation}
with higher $p_{\text{yes}}$ indicating stronger relevance.

\section{Architecture of MixLM}\label{sec:arch}
\begin{figure}
  \includegraphics[width=0.4\textwidth]{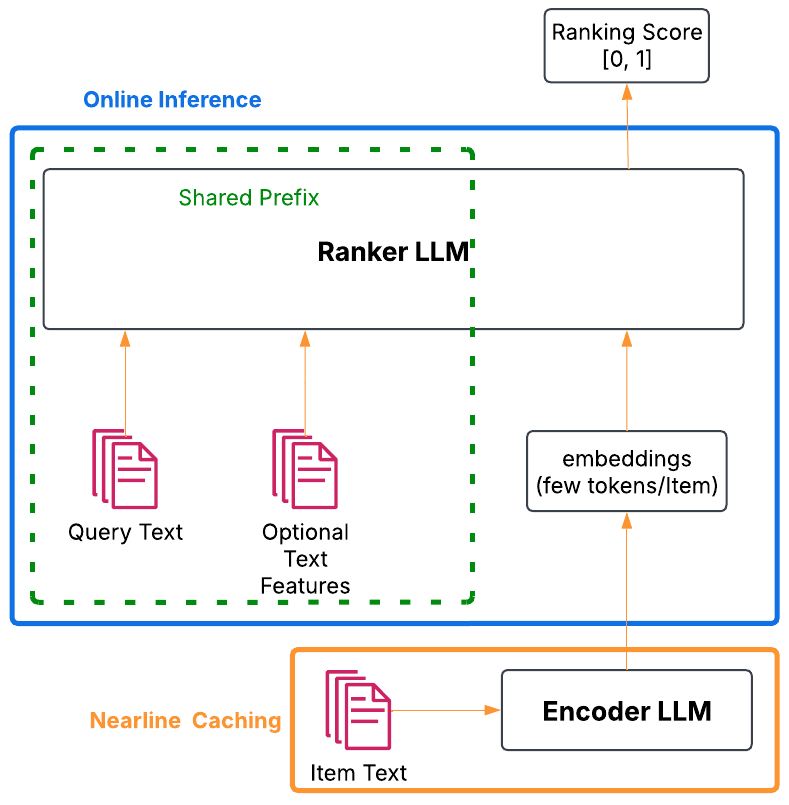}
  \caption{Architecture of MixLM. In the offline stage, item documents are processed by an encoder LLM to obtain compact embedding representations, which are stored in a near-line cache. At serving time, these embeddings are fetched and concatenated with the user’s query and optional auxiliary text features, and the resulting mixed input is passed to the ranker LLM to generate relevance scores.
}
  \label{fig:teaser}
\end{figure}

As the computational complexity of LLM inference scales with input length, reducing the number of tokens can substantially lower inference costs. In this work, we propose a text–embedding mix-interaction architecture that compresses item text description into an embedding vector using an encoder LLM. The resulting embeddings are then combined with other textual inputs and fed into a ranker LLM for output prediction. Because the item embeddings are precomputed and stored in a near-line cache, the number of tokens that must be processed during online inference is greatly reduced, leading to significant efficiency gains. %The detailed design of the ranker architecture and the corresponding training strategies are presented in the following sections.

\paragraph{Detailed Architecture}
Our ranking architecture consists of two components: an encoder LLM, which computes embeddings that summarize item information, and a ranker LLM, which predicts the final task scores. See Figure~\ref{fig:teaser} for an overview.

Before diving deeper into our model architecture, let us 
recall that an LLM can be decomposed into three main components: The input embeddings layer, the transformer blocks and the optional output head. For an LLM, the input embeddings layer creates the input's embedding representation. The transformer layers then form the output embedding representations, given the input embeddings. Finally, the LLM output is generated via an optional output head. For example, for a relevance classification model (in contrast to generative LLMs), we might assume the output head is a binary classification head which generates the estimated probability that the query and item are relevant.

For a sequence of length $T$ and a fixed set of tokenized vocabulary $\mathbb{V}$, we
let $g_R(\cdot; \omega_R): \mathbb{V}^{T} \rightarrow \mathbb{R}^{T \times H}$ 
to denote the input embedding layer of the Ranker LLM, parameterized by $\omega_R$, with a hidden dimension of $H$. We also let $f_R(\cdot; \theta_R): \mathbb{R}^{T \times H} \rightarrow [0,1]$ denote the Ranker LLM decoder and output head, parameterized by $\theta_R$, where the output quantifies the probability of query-item match (i.e., $p_{\text{yes}}$).
 The Ranker LLM can thus be expressed as
\begin{gather}
    F_R(\cdot; \Theta_R) = (f_R \circ g_R)(\cdot; \omega_R, \theta_R): \mathbb{V}^{T} \rightarrow [0,1],
\end{gather}
where $\Theta_R = [\omega_R, \theta_R]$ represents all ranker parameters.  

Similarly, let $\Theta_E = [\omega_E, \theta_E]$ denote the parameters of the Encoder LLM. The Encoder LLM for a sequence of length $T$ can be parameterized as
\begin{gather}
    F_E(\cdot; \Theta_E) = (f_E \circ g_E)(\cdot; \omega_E, \theta_E): \mathbb{V}^{T} \rightarrow \mathbb{R}^{T \times H},
\end{gather}
where $g_E(\cdot; \omega_E): \mathbb{V}^{T} \rightarrow \mathbb{R}^{T \times H}$ is the embedding layer and $f_E(\cdot; \theta_E): \mathbb{R}^{T \times H} \rightarrow \mathbb{R}^{T \times H}$ is the transformer that maps token embeddings to hidden representations (note that we do not have an output head for the encoder model).  

In our mix-interaction system, we decompose the prompt in~\eqref{eq:promt} into two parts, corresponding to ranker and encoder parts, and tokenize each part:
\begin{equation}
    \begin{aligned}
    X_R &= \text{tokenize}([\text{system prefix},q]) \in\mathbb{V}^{T_R} \\
    X_E &= \text{tokenize}([i])\in\mathbb{V}^{T_E}
\end{aligned}
\end{equation}
where $T_R, T_E$ are the lengths of $X_R, X_E$.
Note that $X_E$ does not depend on the query $q$ but only the item description $j$, and therefore, we can encode and store $X_E$ nearline/offline. We then let $h_E = f_E(X_E) \in \mathbb{R}^{T_E \times H}$ to be the encoded item. To compress this item representation, we sample $h_E$ to a shorter sequence of length $T_S$ (specifically in MixLM, we use last N sampling), yielding $h_S = \mathrm{Samp}(f_E(X_E)) \in \mathbb{R}^{T_S \times H}$. In general, one can consider several sampling techniques. In our work, for a give $T_S\geq 1$, we sample the $T_S$ embeddings in $h_E$ corresponding to the last $T_S$ input tokens. In particular, if $T_S=1$, we only take the embedding representation of the last input token.

Next, the input token embeddings of the prefix $X_R$ are computed via the ranker embedding layer, i.e., $h_R = g_R(X_R)$. The two representations $h_R,h_S$ are concatenated to form the combined embedding sequence $[h_R; h_S; h_{EOS}] \in \mathbb{R}^{(T_R+T_S+1) \times H}$, followed by a special end of sentence token. The final ``yes'' probability is predicted by the ranker transformer:
\begin{equation}\label{eq:concat}
    p_{\text{yes}}(q,i;\Theta) = f_R\left([h_R; h_S; h_{EOS}]\right)\in [0,1],
\end{equation}
where $\Theta = [\Theta_R, \Theta_E]$ denotes the full set of trainable parameters.  

Since $h_S = \mathrm{Samp}(F_E(X_e))$ can be precomputed and stored in a near-line cache, the effective sequence length during online inference is reduced from to $T_R + T_S + 1$. In practice, $T_S\ll T_E$, where we might take $T_S=1$ and $T_E$ is 2100 in p99, representing orders of magnitudes of reduction in the prompt length. Furthermore, when ranking multiple items, $X_R$ is shared across all prompts that correspond to the same user and query. As the result, the Key–Value (KV) cache of $X_R$ of be used leading to further reduction of computational complexity.

\section{Training}\label{sec:arch}
In this section, we describe the models, datasets, and training pipeline used to deploy the
\textbf{MixLM} framework in LinkedIn’s semantic job search system. MixLM requires jointly
training a ranker LLM and an encoder LLM, with the key requirement that the encoder’s output
embeddings align with the ranker’s input embedding space so that encoded item representations
can be consumed directly by the ranker (via concatenation in Eq.~\eqref{eq:concat}). As long as
the ranker’s hidden size matches the encoder’s output dimension, the two models may differ
architecturally. For simplicity and training efficiency, however, we use the same
0.6B-parameter architecture for both the ranker and encoder.

We adopt a three-stage training recipe for MixLM, summarized in
Table~\ref{tab:training}. In Stage~I, we perform supervised fine-tuning of a 0.6B pretrained
model on a domain-specific dataset; the resulting checkpoint serves as the base initialization
for the ranker. In Stage~II, we train a pure-text ranker using full item descriptions and
labeled ground truth. This model achieves high ranking quality and is treated as a
\emph{teacher} for subsequent distillation. In Stage~III, we co-train the ranker and encoder
using custom loss functions that distill knowledge from the text-based teacher into the
mixed-input architecture. The ranker is initialized from the Stage~I checkpoint, while the
encoder is initialized from a 0.6B General Text Embedding (GTE) model trained with contrastive
learning.

\begin{table*}[h]
\small
  \caption{MixLM Training Stages: In the stage I, we use the chain-of-thought reasoning traces from our in-house relevance judge to pretrain a 0.6B model on the domain-specific data. In stage II, we apply Supervised Fine-Tuning (SFT) on the model from stage I with the text-only ranking dataset, which serves as a teacher model in the next stage. In stage III, we co-train the encoder and ranker LLMs using a ranking dataset. We include several different loss functions in this stage, including the ranking SFT loss, a 
distillation loss and a self-alignment regularization.}
  \label{tab:training}
  \begin{tabular}{lccc}
    \toprule
    Stages & \makecell{Stage I: \\ Domain-specific Fine-Tuning} & \makecell{Stage II: \\Ranking Teacher Training} & \makecell{Stage III: \\Joint Encoder-Ranking Training} \\
    \midrule
    Data       & Reasoning Dataset     &Ranking Dataset &Ranking Dataset \\
    Number of Samples & 180k    & 10.9M & 10.9M\\
    Input Prompt Type & Text  (chain-of-thoughts) & Text & Text + Embedding\\
    Maximum Sequence Length & 2\,048 & 2\,048 & 2\,176 \\
    Trainable Module(s) & Pretrained LLM  & Ranker LLM  & Encoder LLM  \& Ranker LLM \\
    Training Objective & Distillation from relevance judge & SFT  & SFT, Distillation from ranking teacher, self-alignment\\
    \bottomrule
  \end{tabular}
\end{table*}

% \section{Training of MixLM}
\begin{figure}
  \includegraphics[width=0.5\textwidth]{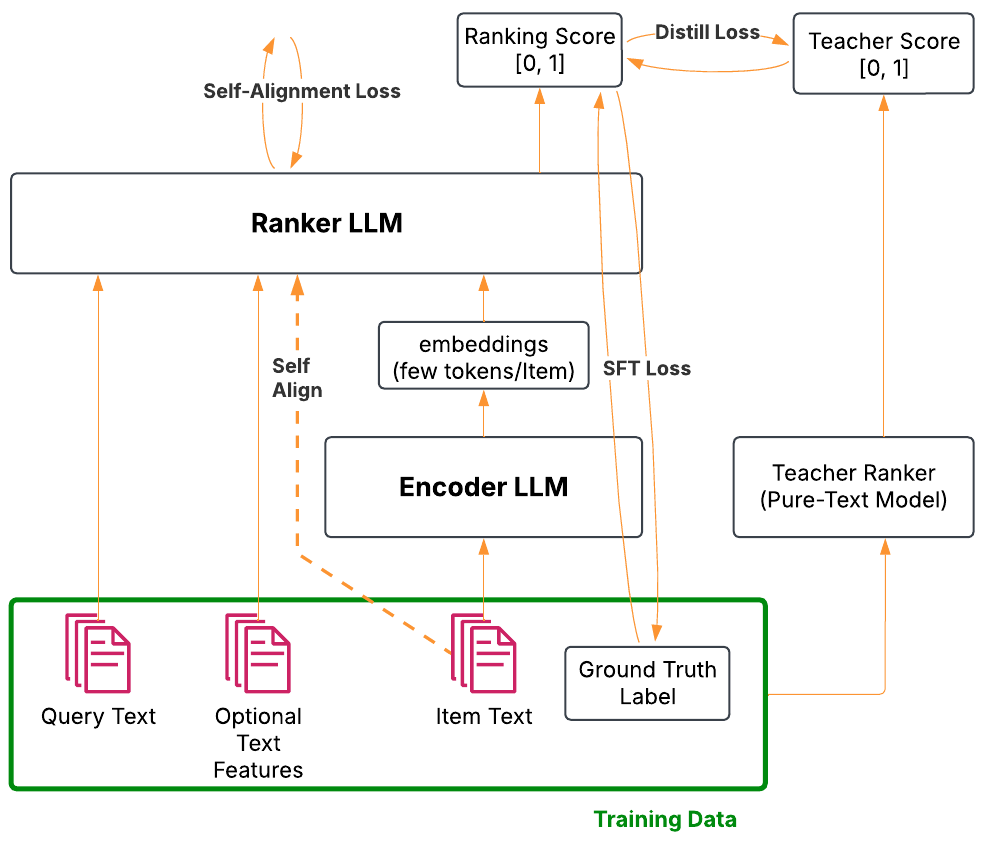}
  \caption{Detailed training pipeline of training state III.}
  \label{fig:loss figure}
\end{figure}

\subsection{Datasets}  
\subsubsection{Label Generation}\label{sec:illuminatorprompt}
We sampled query--item pairs from real user logs, and train a 7B LLM as our relevance judge to generate 5-point graded scores along several dimensions, together with rationales, such that it aligns with the product policy.
The integer labels were subsequently normalized to continuous values in the range $[0, 1]$ to create the ground truth labels $p^*_{\text{yes}}(q,j)$ for query-item pairs.

This 7B in-house LLM operates using the following template:
\begin{verbatim}
[CRITERIA]   : <matching guidelines>
[EXAMPLES]   : <reasoning and output format>
[QUERY]      : <query text>
[CANDIDATE]  : <item text>
Analyze the query--candidate pair and determine whether
they are a good match. Return a single matching score 
(0/1/2/3/4) with detailed reasoning.
\end{verbatim}
We gather model's chain-of-thought traces~\citep{wei2022chain}, as well as the final matching score which we use in different stages of training.

% label  each pair by prompting full-size LLM, e.g. GPT or Claude, 

%   These dimension-wise grades are aggregated into a final label for the pair. 

\subsubsection{Teacher Training Prompts in Stage II}\label{sec:teacher-prompt}
For the textual teacher training, we applied the following chat template to the ranker LLM:
\begin{verbatim}
[SYSTEM]     : Output only 'Yes' or 'No' based on how well 
               the item matches the query; no extra text. 
[QUERY]      : <query text>
[CANDIDATE]  : <item text>
yes or no ?
\end{verbatim}

\subsubsection{Mixed Input Ranking Training Prompts in Stage III}\label{sec:mixedprompt}
For the joint ranking training (co-train the encoder LLM and ranker LLM), we applied the following chat template to the ranker LLM:
\begin{verbatim}
[SYSTEM]     : Output only 'Yes' or 'No' based on how well 
               the item matches the query; no extra text. 
[QUERY]      : <query text>
[CANDIDATE]  : <item embeddings> <EOS>
\end{verbatim}
where the \texttt{<item embeddings>} is the output from encoder LLM. The encoder LLM use the following prompt ``\texttt{Item information: <item text>}''. Note that the ranking prompt used here is significantly shorter compared to the teacher one in Section~\ref{sec:teacher-prompt} which uses full item text.

\subsection{Three-stage Training}
Next, we provide the details of our training pipeline.

\subsubsection{Stage I: Domain Reasoning Fine-Tuning}  
In the first stage, we train a general pretrained LLM using a domain-specific reasoning dataset to strengthen its capability for logical inference and contextual understanding over query–item pairs. This intermediate training phase enables the model to grasp the rationale behind assigning relevance scores, thereby enhancing its generalization performance in downstream ranking tasks.

Concretely, we distill~\citep{hinton2014distilling} our 7B relevance judge into a 0.6B pretrained model, using KL divergence between the logits of the two models. As for the data, we randomly select $180$K samples following the prompt given in Section~\ref{sec:illuminatorprompt}, together with the chain-of-thoughts responses of the relevance judge model.

\subsubsection{Stage II: Ranking Teacher Training}
In the next step, we train a purely textual ranking model. Although this intermediate model cannot be served online due to the long input prompts, it will serve as a teacher model when training the mixed input MixLM model. As we demonstrate, using such a ranking teacher model results in better convergence and training quality.

Specifically, starting from the checkpoint available from our first training stage, we perform Supervised Fine-Tuning (SFT) using the input prompt discussed in Section~\ref{sec:teacher-prompt}. We generate $10.9$M examples based on the prompt provided above, and normalize the integer matching scores, available from the 7B judge, into ground truth probabilities $(p^*_{\text{yes}},p^*_{\text{no}})$. The SFT is then performed using a KL divergence loss. We use $(\hat{p}_{\text{yes}},\hat{p}_{\text{no}})$ to denote the output probability distribution from the final checkpoint for this training stage.

\subsubsection{Stage III: Joint Encoder-Ranking Training}\label{sec:stageiii}
In the final training stage, we train both ranker and encoder models to estimate relevance matching scores, using the mixed input prompt from Section~\ref{sec:mixedprompt}. Specifically, the relevance scores are calculated using~\eqref{eq:concat}. The trainable parameters are  $\Theta = [\Theta_R, \Theta_E]$ which encompasses all parameters of ranker and encoder models.

To this end, we initialize the ranker module in Figure~\ref{fig:teaser} via the final checkpoint from our first stage training. For the encoder module, we use a pretrained 0.6B GTE model. We use the same 10.9M sample dataset from the previous stage for training.
Given the specific structure of the ranking stack, we have observed using custom loss functions can improve the training quality and the final models' ranking quality.  Below we discuss the details of the loss functions we use.

\paragraph{Soft-Label SFT Loss} Our first loss function is a KL divergence SFT loss, that ensures the predicted probabilities from MixLM are close to the ground truth. This loss function is given as
$$\mathcal{L}_{\text{SFT}}(\Theta)=\text{KL}\left((p^*_{\text{yes}},p^*_{\text{no}})||({p}_{\text{yes}}(\Theta),{p}_{\text{no}}(\Theta))\right)$$
where $({p}_{\text{yes}},{p}_{\text{no}})$ are the predicted probability of relevance match (see~\eqref{eq:concat}), and $(p^*_{\text{yes}},p^*_{\text{no}})$ are ground truth probabilities.

\paragraph{Ranking Distillation Loss} In addition to comparing to the ground truth, we compare the output distribution from MixLM to the one from the ranking teacher model from Stage II. More specifically, we define
$$\mathcal{L}_{\text{distill}}(\Theta)=\text{KL}\left((\hat{p}_{\text{yes}},\hat{p}_{\text{no}})||({p}_{\text{yes}}(\Theta),{p}_{\text{no}}(\Theta))\right)$$
where we recall that, $(\hat{p}_{\text{yes}},\hat{p}_{\text{no}})$ is the output distribution of ranking teacher with full text prompt in Section~\ref{sec:teacher-prompt}.

This loss function is inspired by knowledge distillation~\citep{hinton2014distilling}, where knowledge from a (typically larger) teacher model is transferred to a (usually smaller) student model. In contrast, our student model, MixLM, which consists of an encoder and a ranker, can in fact be larger than the teacher model. As discussed in Section~\ref{sec:experiments}, the teacher operating on pure text is easier to train to provide high-quality predictions that are cleaner than the original dataset labels. When training MixLM, the teacher label helps reducing gradient variance and facilitates better convergence.

\paragraph{Self-Alignment Loss} Finally, we use two self-alignment loss functions, that help to ensure the encoder model's output embeddings are aligned with the input embeddings of the ranker model. From our experience, including these loss functions has a regularizaiton effect that can prevent overfitting. To this end, we pass the full text prompt (from Section~\ref{sec:teacher-prompt}) through the mixed input model's ranker module (the model parameterized by $\Theta_R$ from section~\ref{sec:arch}), and collect the output probability distribution (which we denote by $(\bar{p}_{\text{yes}},\bar{p}_{\text{no}})$). Additionally, we collect the last layer's (before the output head) hidden states corresponding to the last input token, which we denote by $\bar{h}_{\text{last}}\in\mathbb{R}^{H}$. Note that these quantities are functions of only $\Theta_R$. For the same sample, we also collect the output probability distribution (denoted as $(p_{\text{yes}},p_{\text{no}})$) and  the last layer's hidden states for the last input token (denoted as $h_{\text{last}}\in\mathbb{R}^H$) when we use the mixed input prompt from Section~\ref{sec:mixedprompt}. These quantities, as expected, are functions of all trainable parameters $\Theta$. Then, we use loss functions that encourage the resulting quantities from these two setups to be similar. First, we use a loss function $\mathcal{L}_{\text{hidden-align}}$ to encourage the last hidden states from the mixed input and textual prompts to be similar:
$$\mathcal{L}_{\text{hidden-align}}(\Theta) = 1 - \mathrm{cossim} \big(h_{\text{last}}(\Theta), \bar{h}_{\text{last}}(\Theta_R)\big) $$
where $\mathrm{cossim}$ denotes the cosine similarity of two vectors. 
In addition, we use another alignment loss to ensure the predicted probabilities remain similar:
$$\mathcal{L}_{\text{pred-align}}(\Theta) = \text{KL}\left((\bar{p}_{\text{yes}}(\Theta_R),\bar{p}_{\text{no}}(\Theta_R))|| (p_{\text{yes}}(\Theta),p_{\text{no}}(\Theta))\right). $$
The overall alignment loss is a linear combination of hidden state and prediction alignment losses:
$$\mathcal{L}_{\text{align}}(\Theta) = \alpha \cdot \mathcal{L}_{\text{pred-align}}(\Theta) \;+\; \beta \cdot \mathcal{L}_{\text{hidden-align}}(\Theta)$$
where $\alpha,\beta\geq 0$ are given hyper-parameters.

Finally, the full training objective integrates all components discussed above. Specifically, we use
$$\mathcal{L}_{\text{total}} (\Theta) = \mathcal{L}_{\text{SFT}}(\Theta) 
+ \lambda_{\text{distill}} \mathcal{L}_{\text{distill}}(\Theta)
+ \lambda_{\text{align}} \mathcal{L}_{\text{align}}(\Theta),
$$
where $\lambda_{\text{distill}}\geq0$ and $\lambda_{\text{align}}\geq 0$ are hyper-parameters that are given a priori. Figure~\ref{fig:loss figure} summarizes the training pipeline for this stage.

% \begin{remark}
%     We note that ranking distillation loss and self-alignment loss functions follow different objectives. The distillation loss provides training guidance from a purely text-based ranking stack, that was trained without using any mixed input prompt. On the other hand, the self-alignment ensure MixLM's ranker module remains aligned, regardless if it is given a mixed input prompt or a text prompt. In particular, the self-alignment loss does not depend at all on the ranking teacher model from the second stage.
% \end{remark}

% Like multi-modal language models, such as LLaVA~\cite{liu2023visual} and BLIP-2~\cite{li2023blip}, the goal of Stage~III is to align the input space of the ranker LLM and the embedding space of the encoder LLM. This is challenging because 1) unlike the discrete vocabulary of text, item embeddings have an infinitely large continuous space, resulting in higher sample complexity; 2) the embeddings encode much more complex and entangled information than text tokens, requiring greater training effort to preserve and decode such information; and 3) item embeddings have never been encountered during foundational pretraining.

To train  MixLM more effectively, we apply a phased curriculum learning strategy, decomposing the joint optimization into two distinct phases with different loss weights, i.e., ($\lambda_{\text{distill}}$ and $\lambda_{\text{align}}$): In \textbf{Phase 1}, we prioritizes the space-alignment by increasing $\lambda_{\text{align}}$ and decreasing $\lambda_{\text{distill}}$, so that we can train the encoder to produce embeddings that integrate seamlessly with the ranker's internal representations. In \textbf{Phase 2}, decrease $\lambda_{\text{align}}$ and increase $\lambda_{\text{distill}}$, focusing on adjusting ranker LLM to improve the relevance score prediction. 

% This approach mirrors the multi-stage training strategies successfully employed in multi-modal LMs, where phased training with different optimization objectives in each phase improves both alignment quality and downstream task performance. During the training, we employ a standard AdamW optimizer, combined with a cosine annealing scheduler with linear warmup.

\input{inference_infra}

\input{experiments}

\input{related_work}

\section{Conclusion}
In this work, we introduced MixLM, a novel framework that bridges the gap between the semantic richness of large language model–based rankers and the stringent efficiency requirements of industrial search and recommender systems. By representing candidate items as compact embedding tokens and introducing the mix-interaction mechanism, MixLM effectively reduces the input context length while maintaining the expressiveness of full-text LLMs. Our co-training strategy for online rankers and near-line embedding flows further enables tight coupling between semantic understanding and serving efficiency.

Empirical results from deployment in a large-scale professional social network demonstrate that MixLM delivers substantial system gains—10.0x times thorughput compared with summarized-text LLM, 75.9x times thorughput compared with full-text LLM, under same latency budget. MixLM enabled LLM search system's deployment on all LinkedIn users' traffic, thereby producing a measurable 0.47\% increase in daily active users (DAU). These findings highlight the potential of mixed-input architectures to reshape the design space of LLM-based ranking systems, offering a practical path toward scalable, low-latency, and semantically powerful search in real-world environments.

Future work will explore adapting co-trained embeddings for modeling member history and exploring MixLM for retrieval tasks.

\bibliographystyle{ACM-Reference-Format}
\bibliography{sample-base}

\section{APPENDIX}
\subsection{Additional Ablation Analysis}
\subsubsection{Curriculum Learning Strategy}

In addition to the auxiliary loss components, we investigated the effect of different phased training strategies that align with curriculum learning principles for Stage III in Section~\ref{sec:stageiii}. Curriculum learning suggests that training with carefully sequenced objectives or loss weightings can lead to better convergence and final model performance. In the context of text-embedding alignment problem, we explored the following curriculum strategies:

\begin{itemize}
\item \textbf{No Curriculum Learning (Baseline)}: focus more on the KL divergence loss but less on distillation loss and self-alignment regularization.

\item \textbf{Two-Phase Curriculum Strategy}: On top of the baseline, add an alignment phase at the beginning that has high self-alignment weights with low KL divergence weights.

\item \textbf{Three-Phase Curriculum Strategy}: Add an intermediate phase to two-phase curriculum, that has equal weighting across all three loss objectives (distillation, SFT, and self-alignment).
\end{itemize}

As seen in Table~\ref{tab:curriculum-learning}, curriculum learning improves ranking performance, indicating the alignment is necessary. The two-phase curriculum outperforms the three-phase approach, suggesting that a direct transition between alignment-focused and task-performance-focused training is more effective than including an intermediate balanced phase. This finding indicates that once the ranker-encoder alignment is sufficiently established, the model can more directly optimize for the downstream ranking task without the gradual intermediate phase.

\begin{table}[h]
    \caption{Ablation study on curriculum learning strategies, on small sub-dataset}
    \label{tab:curriculum-learning}
    \begin{tabular}{l c}
    \hline
    \textbf{Curriculum Strategy} & \textbf{NDCG@10} \\
    \hline
    no curriculum learning (Task)                    & -- \\
    two-phase (Alignment $\to$ Task)          & \textbf{+0.0020} \\
    three-phase (Alignment $\to$ Balanced $\to$ Task) & +0.0015 \\
    \hline
    \end{tabular}
\end{table}

\subsection{Training Efficiency}
During the ranking training phase, we co-train the encoder LLM and the ranker LLM end-to-end. We deploy PyTorch’s Fully Sharded Data Parallel~\citep{zhao2023fsdp} to shard the decoder parameters, gradients, and optimizer states across 8 nodes (each with 8 NVIDIA H100 GPUs). Within each node, GPUs communicate via NVLink, and inter-node communication is via InfiniBand.

To reduce communication overhead, we enable layer-wise prefetching: before computing the next forward layer, we prefetch its parameters; during the backward pass, we prefetch gradients of upcoming layers. To further reduce memory footprint, we incorporate Liger Kernel~\citep{hsu2024liger}, a set of optimized Triton kernels (e.g. fused RMSNorm, RoPE, SwiGLU, fused CrossEntropy) that deliver higher throughput and lower memory usage.

In our setup, the total compute cost to train on 70B tokens under the ranking objective is approximately 700 H100 GPU-hours.

\begin{figure}
    \centering
    \includegraphics[width=1\linewidth]{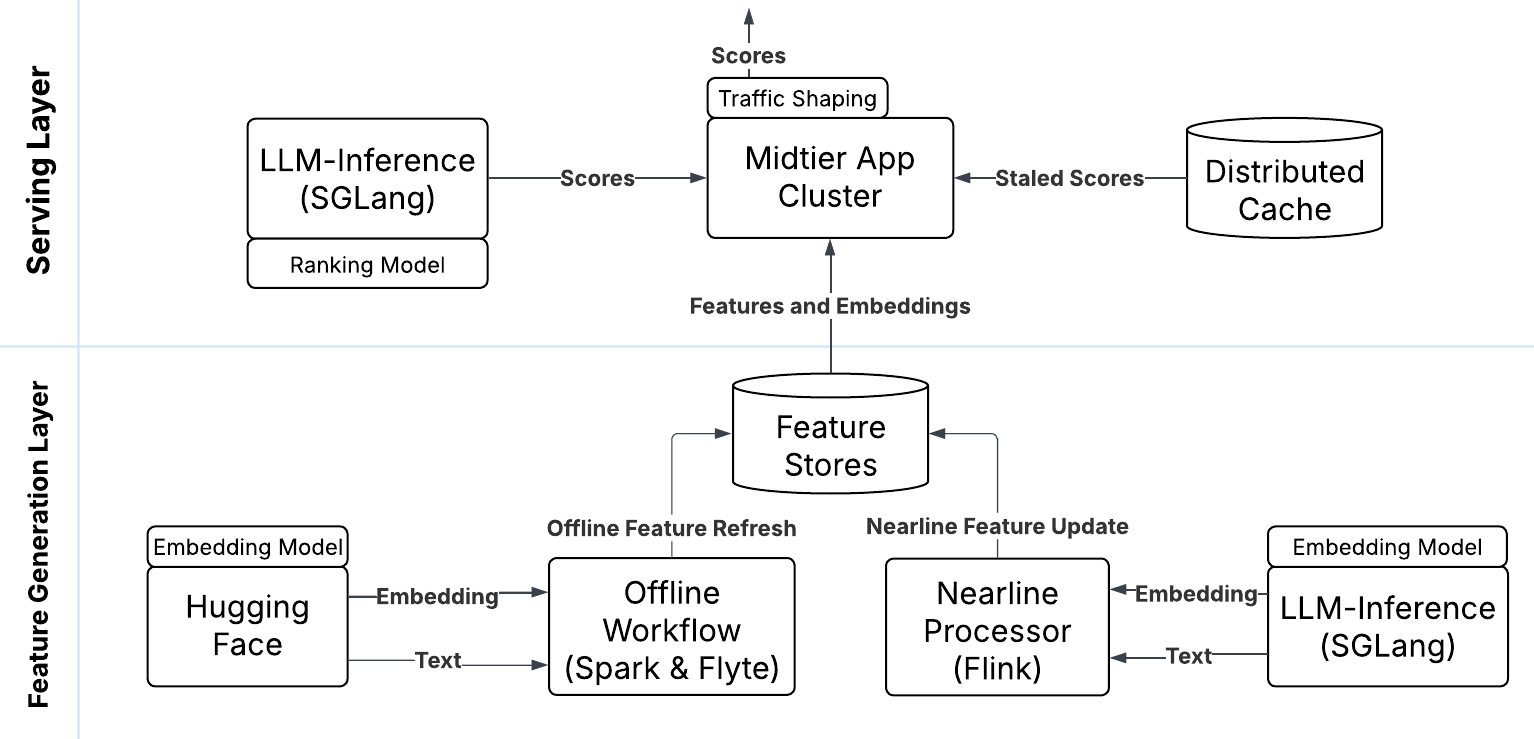}
    \caption{Online Serving Architecture for LinkedIn's Semantic Job Search}
    \label{fig:serving}
\end{figure}

\begin{figure}[h]
  \centering
  \includegraphics[width=0.4\textwidth]{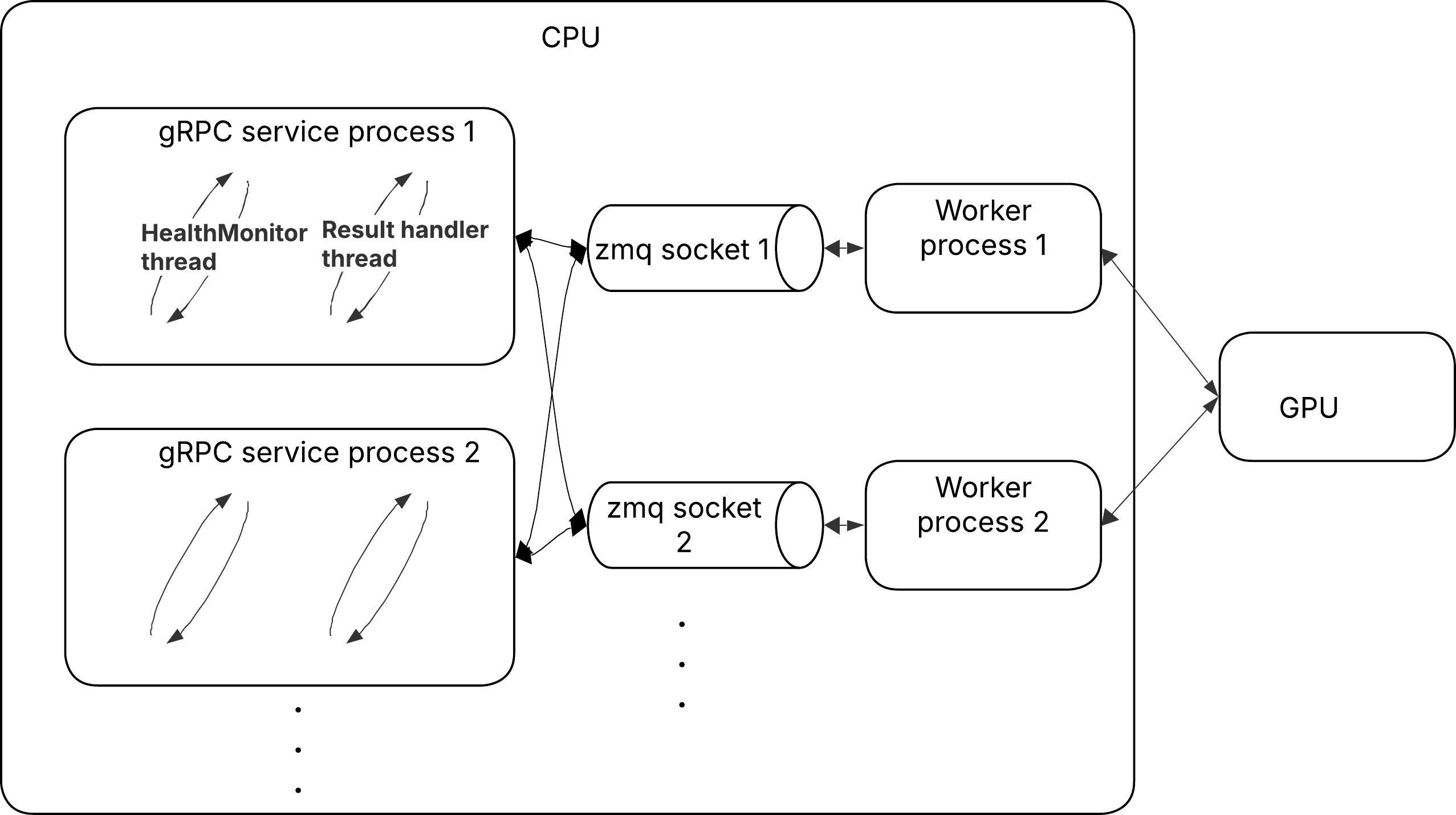}
  \caption{Multiprocessing server architecture for gRPC}
  \label{fig:multi_server}
\end{figure}

\subsection{Online System Implementation}

The online inference architecture adopts a two-layer design consisting of a feature generation layer and a serving layer (see Figure~\ref{fig:serving} for an overview). The feature generation layer functions as a hybrid system that combines offline refresh with near-line ingestion and generation to maintain comprehensive and up-to-date feature coverage. The offline pipeline, orchestrated through Flyte, periodically performs large-scale GPU-based inference to regenerate the full corpus of text and embedding features. In parallel, the near-line component continuously ingests real-time updates reflecting changes in the population of scoring candidates and generates corresponding features through GPU inference. Because near-line processing is not latency-sensitive, GPU resources can be utilized with high throughput and efficiency. All generated features are persistent in a high-throughput distributed storage system, ensuring consistency, low-latency availability for downstream serving.

The serving layer performs real-time inference by retrieving precomputed features and assembling mixed text–embedding prompts that integrate user queries with candidate scoring items. Prior to dispatching requests to the SGLang inference engine~\citep{sglang} for GPU-based scoring, several system-level optimizations—including distributed caching, latency reduction mechanisms, and scalable serving infrastructure—are applied to minimize feature retrieval overhead and improve serving efficiency. This layered architecture enables effective reuse of precomputed representations while preserving responsiveness across both near-line and real-time applications. Empirical evaluations demonstrate that these design choices significantly enhance inference efficiency, GPU utilization, and system scalability, supporting production-scale deployment of large language model–based ranking systems.

\subsection{LinkedIn Semantic Job Search}In this work, we focus on LinkedIn’s Semantic Job Search, which redefines job discovery through an AI-driven focus on intent rather than keywords. By interpreting natural language inputs, it captures users’ underlying goals and preferences, minimizing dependence on exact word matches between queries and job descriptions. This semantic understanding allows for more accurate and flexible retrieval, overcoming vocabulary gaps and delivering search results that align more closely with how members naturally express their career interests.

In practice, the prompt in~\eqref{eq:promt} can be lengthy. In fact, in our semantic job search system, most of the prompt will be dominated with item description $j$, which has a median length of 900 tokens, and can take up to 2100 tokens. The lengthy prompts lead to  unacceptable latency and throughput overhead, which makes deploying such a text-based system to production costly.
To handle job search traffic spanning various product surfaces, our ranking system must process and score about 3.15 million items per second, making ranking efficiency absolutely critical.

Even though through previous summarization-solution\citep{behdin2025scaling} we were able to reduce the item description length and rollout LLM ranking in limited-scale traffic, the prompt context length remains the major computational bottleneck and hard blocker for LinkedIn's Semantic Job Search's full deployment. 

\subsection{LLMs for Recommender Systems}
Recent years have seen a surge of interest in utilizing LLMs for search and recommender system applications. For example, several papers have reported that LLMs' understanding of natural language makes them powerful relevance judges. Moreover, LLMs are strong ranker which can extract users' preferences based on their past interactions~\citep{relevance4}. This has prompted a long line of research, investigating how LLMs can be integrated to existing recommender system, or even used as standalone recommenders~\citep{llmrs}. For example,~\citep{related1} show that LLMs can be successful in solving the cold-start problem in recommender systems, where not much historical data is available for the user. The cold-start problem was further studied by~\citep{related5} through the lens of instruction tuning. A similar approach was also used by~\citep{related4}.~\citep{related3} discuss using LLMs to predict user actions and preferences when the model is presented with the past data as a text prompt.

\subsection{Industrial Application of LLMs for RecSys}
In addition to understanding the methodological aspects of MixLM, we also focus on a real-world semantic search that is deployed at an industrial scale. Related to this,~\citep{related-pinterest} discuss how an LLM can be used as a relevance judge. However, the LLM-based system is eventually distilled into a non-LLM architecture for online serving. Similar challenges have been reported by~\citep{related-ebay} and~\citep{related-walmart}, where either the model or some of the text features need to be altered for online serving. This further motivates our work to study methods that improve the efficiency of LLM-based recommender systems. In a recent work,~\citep{relevance-tencent} study the distillation an LLM-based system into a BERT-type~\citep{bert} model. Notably, for most these real-world applications discussed here, to meet efficiency goal, either a non-LLM model has to be deployed, or some features have to be removed. In contrast, in our real-world application, we deploy the LLM online, and all features are consumed by our encoder model to create item embedding representations.

\end{document}

%% file: introduction.tex
\section{Introduction}

Large Language Models (LLMs) have been widely adopted in industrial applications due to their
strong generative and predictive capabilities. In search and recommender systems, LLMs
deliver substantial relevance improvements by capturing subtle linguistic cues and conceptual
relationships that traditional lexical methods often miss
\citep{relevance2,relevance3,relevance4,llmrs}. When used as relevance judges or rankers, LLMs
enable semantic retrieval and ranking that better aligns with user intent.

Despite these strengths, deploying LLMs in industrial search and recommendation systems
remains challenging due to their high computational cost, particularly when processing long
text inputs under strict latency and throughput constraints. A common cross-encoder approach
forms prompts that combine user queries with candidate item descriptions and asks the LLM to
assess relevance \citep{relevance2,relevance3,relevance-biencoder,rankvicuna}. Such prompts
often contain thousands of tokens, resulting in heavy \emph{prefill}-dominated inference and
quadratic attention costs with respect to context length. In practice, this forces systems to
either replace LLMs with smaller models or drop informative features from online serving
pipelines \citep{related-ebay,related-pinterest,related-walmart}. Reducing input context
length while preserving semantic expressiveness is therefore a central challenge for
LLM-based ranking.

In this work, we introduce \textbf{MixLM}, an LLM-based ranking framework designed to
substantially reduce input context length while retaining the semantic richness of full-text
LLM rankers. MixLM compresses each candidate document—often thousands of tokens long—into a
small set of learned embedding tokens using an encoder LLM. These embeddings are stored in a
nearline cache and, at inference time, are mixed with the user’s natural-language query and
auxiliary text and processed by a ranker LLM. By replacing long document text with a handful
of embedding tokens, MixLM reduces computational complexity by orders of magnitude while
preserving rich query–document interactions.

Beyond the modeling formulation, we describe the end-to-end training and deployment of MixLM
in LinkedIn’s production semantic job search system. We introduce a joint training pipeline
that optimizes both encoder and ranker LLMs and propose distillation losses that transfer
knowledge from a strong full-text LLM ranker into the mixed-interaction architecture. We also
analyze scaling behavior with respect to training data size and the number of embedding tokens
per item.

From a systems perspective, we present a detailed analysis of MixLM’s impact on serving
infrastructure. Context compression via embedding tokens enables substantially higher ranking
throughput, which is further amplified by nearline caching and amortized prefill computation.
Together, these optimizations compound to deliver large efficiency gains for mixed-input LLM
serving.

We validate MixLM at industrial scale through deployment in LinkedIn’s AI-driven Job Search.
Under the same latency budget, MixLM:
\begin{itemize}\setlength\itemsep{0pt}
    \item improves throughput by \textbf{10.0$\times$} compared to summarized-text LLM ranking,
    \item improves throughput by \textbf{75.9$\times$} compared to full-text LLM ranking, and
    \item preserves the relevance quality of a strong full-text LLM baseline.
\end{itemize}
These efficiency gains enabled full-traffic deployment of LLM-based ranking and resulted in a
\textbf{0.47\%} lift in Daily Active Users (DAU) in a large-scale online A/B experiment.

\paragraph{Summary of Contributions.}
\begin{itemize}\setlength\itemsep{0pt}
\item We introduce \textbf{MixLM}, a mixed-input LLM ranking framework that replaces long item
text with compact embedding tokens to reduce prompt length.
\item We propose an end-to-end training pipeline with distillation losses that align mixed-input
ranking with full-text LLM behavior.
\item We present a production-ready serving stack for mixed-input LLM ranking, achieving
order-of-magnitude improvements in GPU throughput.
\item We share practical lessons from deploying MixLM in LinkedIn’s semantic job search at full
production scale.
\end{itemize}

Together, these contributions demonstrate a practical and scalable path for deploying
LLM-based ranking systems that combine deep semantic understanding with real-time efficiency.

%% file: inference_infra.tex
\section{Inference Engine Optimization}

\subsection{Text--Embedding Mixed Input}
% Most open-source LLM inference engines assume pure-text inputs, limiting requests to tokenized
% sequences. 
To support MixLM’s mixed-interaction design, we extend the serving interface to
accept precomputed embedding tensors alongside text. Embeddings are transmitted via a
structured \emph{feature payload} added to the gRPC request schema, consisting of a binary
tensor buffer and metadata specifying shape, data type, and optional serialization format
(e.g., NumPy~\citep{numpy}).

On the server side, the preprocessing module decodes and validates feature payloads into native
tensors using metadata-driven reshaping and casting, and attaches them to the model input
context. This enables direct reuse of embeddings produced by upstream pipelines, avoiding
redundant computation and reducing latency. The schema remains fully backward-compatible with
text-only requests and introduces minimal serialization overhead.

\begin{table*}[t]
\small
  \caption{Attention and linear-layer compute under different conditions, in terms of $(T_q,
  T_i, N_i)$, where $T_i$ denotes full item-text length and $K$ is the MixLM compression factor
  (item length becomes $T_i/K$). $\propto$ denotes proportionality up to
  architecture-dependent constants. Bottom blocks instantiate $N_i = 250$, $K = 450$.}
  \label{tab:complex1}
  \begin{tabular}{lll}
    \toprule
    \textbf{Conditions} 
    & \textbf{Total attention cost} 
    & \textbf{Total linear-layer cost} \\
    \midrule
    Naive (full-text items)
    & $\propto N_i (T_q + T_i)^2$
    & $\propto N_i (T_q + T_i)$ \\
    \hline
    Amortized prefill (full-text items)
    & $\propto T_q^2 + N_i (2 T_i T_q + T_i^2)$
    & $\propto T_q + N_i T_i$ \\
    \hline
    \makecell[l]{Amortized prefill \\ $+$ MixLM}
    & $\propto T_q^2 + N_i \bigl( 2 (T_i/K) T_q + (T_i/K)^2 \bigr)$
    & $\propto T_q + N_i T_i / K$ \\
    \midrule
    \makecell[l]{Naive, $N_i=250$}
    & $\propto 250 (T_q + T_i)^2$
    & $\propto 250 (T_q + T_i)$ \\
    \hline
    \makecell[l]{Amortized, $N_i=250$}
    & $T_q^2 + 500 T_i T_q + 250 T_i^2$
    & $\propto T_q + 250 T_i$ \\
    \hline
    \makecell[l]{Amortized $+$ MixLM \\ $N_i=250$, $K=450$}
    & $T_q^2 + \frac{10}{9} T_i T_q + \frac{1}{810} T_i^2$
    & $T_q + \frac{5}{9} T_i$ \\
    \bottomrule
  \end{tabular}
\end{table*}

\subsection{Shared-Prefix Prefill Optimization}

In large-scale ranking, each request typically scores hundreds to thousands of items sharing
the same query and user context. After MixLM compresses each item into a small number of
embedding tokens (from a median of $\sim$900 tokens to two tokens in production), inference
cost becomes dominated by repeated query-prefix computation.

To improve GPU utilization, we batch candidate-item prompts per query and reuse the KV cache
across prompts with a shared prefix. This is particularly effective under MixLM, where item
tokens are minimal and query-prefix computation dominates cost.

Let $T_q$ be the query-prefix length, $T_i$ the full-text item length, and $N_i$ the ranking
depth. Ignoring architecture-dependent constants, attention and linear costs scale as
\[
\text{Attention} \propto L^2, \qquad \text{Linear} \propto L.
\]

\paragraph{Naïve prefill (full-text).}
Each item is processed independently with length $(T_q + T_i)$:
\[
F_{\text{att}}^{\text{naive}} \propto N_i (T_q + T_i)^2, \qquad
F_{\text{lin}}^{\text{naive}} \propto N_i (T_q + T_i).
\]

\paragraph{Amortized prefill (full-text).}
Query-prefix attention is computed once and reused:
\[
F_{\text{att}}^{\text{amort, full}}
\propto
T_q^2 + N_i (2 T_i T_q + T_i^2),
\qquad
F_{\text{lin}}^{\text{amort, full}} \propto T_q + N_i T_i.
\]

\paragraph{Amortized prefill with MixLM.}
With item compression factor $K$, each item contributes $T_i/K$ tokens:
\[
F_{\text{att}}^{\text{amort, MixLM}}
\propto
T_q^2 + N_i \bigl( 2 (T_i/K) T_q + (T_i/K)^2 \bigr),
\]
\[
F_{\text{lin}}^{\text{amort, MixLM}} \propto T_q + N_i T_i / K.
\]
Thus, item-side attention and linear costs are reduced by approximately $K$.

We implement amortization using two mechanisms:
\begin{itemize}
    \item \textbf{In-batch prefix caching:} We reuse the key--value (KV) states from the first prompt so that all items in the batch share the same query-prefix computation in a single forward pass. The attention computation merges two operations: (1)suffix tokens attend to all prefix tokens from the first prompt via dense paged attention; (2)suffix tokens attend to themselves via regular causal attention.
    \item \textbf{Multi-item scoring:} Multiple items are concatenated into a single sequence
    separated by delimiters \cite{sglang_pr_10979}. FlashInfer applies item-aware masking to
    prevent cross-item attention.
\end{itemize}

Because MixLM reduces $T_i$ to only a few tokens, shared-prefix amortization eliminates most
redundant computation and delivers substantial inference-throughput gains in production.

\subsection{Inference Engine CPU Overhead Reduction}

\subsubsection{Multi-Process gRPC}
After optimizing the model and the SGLang inference engine, request preprocessing in the Python
gRPC service emerged as the dominant bottleneck. Although the gRPC server uses asynchronous
Python, request deserialization, feature decoding, and context construction are constrained by
the Global Interpreter Lock (GIL) and the single-threaded asyncio event loop. Under high request
rates, event-loop contention and per-request overhead limit throughput despite asynchronous
I/O.

To address this, we adopt a multiprocessing server architecture that decouples the gRPC
frontend from the SGLang engine. Request handling and preprocessing are parallelized across
multiple CPU processes, while the SGLang engine runs independently. This design fully utilizes
multi-core CPUs and improves throughput by 40\%.
\subsubsection{Batch Send}
To fully exploit amortized prefill, the SGLang scheduler must admit prompts sharing the same
query as a single batch. We therefore introduce an explicit \emph{batch send} mechanism, where
all item prompts for a query are serialized and transmitted as a single ZMQ message. This
ensures that the scheduler processes the batch atomically and can reliably apply in-batch
prefix caching.

\subsubsection{Multi-Process Parallelization}
While GPU prefill executes efficiently, request scheduling on a single CPU core became a
secondary bottleneck. We mitigate this by running multiple scheduler processes (e.g., five)
per GPU, each assigned a dedicated GPU memory partition (e.g., 20\% per worker). This
parallelizes CPU-side scheduling while maintaining high GPU utilization, effectively
pipelining batch preparation with GPU execution.

\subsubsection{Eliminating Redundant KV-Cache Operations}
SGLang’s default KV-cache behavior is designed for autoregressive decoding, which is not used
in our prefill-only scoring workload. We modify cache management to immediately release KV
cache memory after request completion, eliminating unnecessary allocation, eviction, and pool
management. This both reduces CPU overhead on the critical path and increases effective GPU
memory capacity, enabling larger batch sizes and higher concurrent throughput.

%% file: experiments.tex
\section{Experimental Results}\label{sec:experiments}

We conduct both offline and online experiments to evaluate the ranking quality and serving
efficiency of MixLM. Offline evaluation uses a held-out test set labeled by our internal LLM
relevance judge. We compare against several baselines: (1) an embedding-based bi-encoder
retrieval model distilled from the LLM judge (the first-stage retriever in our stack),
(2) a pure-text 0.6B LLM ranker from the second training stage, representing an upper-bound
model that is impractical to deploy, and (3) LinkedIn’s limited-scale production baseline
that applies near-line text summarization and aggressive ranker pruning
\citep{behdin2025scaling}.

We report NDCG@10 for ranking quality and maximum throughput measured as items scored per GPU
per second, under a fixed p99 end-to-end latency constraint of 500 ms.

\subsection{Efficiency and Relevance}

\begin{table}[t]
  \caption{Ranking quality comparison of several ranking approaches. We report NDCG@10 with
  respect to labels provided by our relevance judge.}
  \label{tab:sota1}
  \small
  \begin{tabular}{lc}
    \toprule
    Model & NDCG@10\\
    \midrule
    Full Text   & 0.9432\\
    Summarized, Pruned & 0.9218\\
    \textbf{MixLM}  & 0.9239\\
    \midrule
    Embedding Retrieval  & 0.8380\\
    \bottomrule
  \end{tabular}
\end{table}

\begin{table}[t]
  \caption{GPU efficiency comparison of ranking approaches. Throughput is measured as maximum
  items scored per GPU per second under a 500 ms latency budget.}
  \label{tab:sota2}
  \small
  \begin{tabular}{lrrr}
    \toprule
    Model & QPS (Items/s/GPU) & Latency (ms) & GPU\\
    \midrule
    Full Text      & 290 & $<$500 & H100\\
    Summarized, Pruned & 2{,}200 & $<$500 & H100\\
    \textbf{MixLM} & 22{,}000 & $<$500 & H100\\
    \midrule
    Embedding Retrieval & $>1.6\times10^9$ & $<$100 & A100\\
    \bottomrule
  \end{tabular}
\end{table}

Table~\ref{tab:sota1} shows that embedding-only retrieval delivers the lowest ranking quality
and is therefore suitable only as a coarse first-stage filter. Full-text LLM ranking achieves
the highest NDCG@10 (0.9432) but is prohibitively expensive, scoring only 290 items/s/GPU.
Summarized-text LLM ranking with pruning improves throughput to 2,200 items/s/GPU while
sacrificing some quality (NDCG@10 = 0.9218).

MixLM achieves comparable ranking quality (NDCG@10 = 0.9239) while scaling throughput to
22,000 items/s/GPU—representing a 10.0$\times$ speedup over summarized-text LLM ranking and a
75.9$\times$ speedup over full-text LLM ranking—at a modest 1.8-point NDCG gap relative to the
full-text model. This places MixLM at a favorable point in the quality–efficiency trade-off,
making it suitable for latency-critical production deployment.

\subsection{Online A/B Test}

\begin{table}[t]
    \caption{Online A/B Test Results}
  \label{tab:abtest}
  \small
  \begin{tabular}{lc}
    \toprule
    Experiment Group & $\Delta$DAU \\
    \midrule
    Classic Job Search & -- \\
    \textbf{Semantic Job Search (MixLM)} & \textbf{+0.47\%} \\
    \bottomrule
  \end{tabular}
\end{table}

Table \ref{tab:abtest} summarizes the findings. After deploying MixLM in LinkedIn’s AI Job Search and running extensive A/B tests, we observed: MixLM has on-par relevance metrics comparing with the existing limited-scale production baseline(summarized text LLM) ~\citep{behdin2025scaling}. With the 10.0x gain in serving throughput under the same latency budget, MixLM empowered the full-traffic deployment of LLM-ranking-based Semantic Job Search and for the first time, consequently producing a significant 0.47\% increase in Daily Active Users, compared with the traditional job search.

\subsection{Ablation Analysis}

To understand the contribution of individual components, we perform ablation studies on a
smaller dataset that mirrors the distribution of the full training corpus.

\subsubsection{Scaling Up}

\begin{table}[t]
  \caption{Ablation on Dataset Size}
  \label{tab:ablationsample}
  \small
  \begin{tabular}{lc}
    \toprule
    Training Samples & $\Delta$NDCG@10 \\
    \midrule
    160K & -- \\
    400K & +0.0250 \\
    800K & +0.0280 \\
    1.08M & +0.0334 \\
    \bottomrule
  \end{tabular}
\end{table}

\begin{table}[t]
  \caption{Embedding Tokens per Item}
  \label{tab:ablationtoken}
  \small
  \begin{tabular}{lc}
    \toprule
    Tokens / Item & $\Delta$NDCG@10 \\
    \midrule
    1 & -- \\
    5 & +0.0017 \\
    10 & +0.0044 \\
    20 & +0.0111 \\
    30 & +0.0158 \\
    40 & +0.0172 \\
    50 & +0.0198 \\
    \bottomrule
  \end{tabular}
\end{table}

Increasing both training data size and embedding granularity consistently improves ranking
quality. For production, we retain a single embedding token per item to meet latency
constraints.

\subsubsection{Domain Reasoning Fine-Tuning}

\begin{table}[h]
\caption{Effect of Domain Reasoning Fine-Tuning}
\label{tab:ranker-quality}
\small
\centering
\begin{tabular}{lc}
\toprule
Ranker Base Model & $\Delta$NDCG@10 \\
\midrule
Vanilla LLM & -- \\
Domain-Reasoning Tuned LLM & \textbf{+0.0185} \\
\bottomrule
\end{tabular}
\end{table}

Domain reasoning fine-tuning substantially improves ranking quality and is adopted as the
first stage of our training pipeline.

\subsubsection{Self-Alignment and Auxiliary Losses}

\begin{table}[h]
\caption{Ablation on Auxiliary Losses}
\label{tab:ablationloss}
\small
\begin{tabular}{lc}
\toprule
Setup & $\Delta$NDCG@10 \\
\midrule
No auxiliary loss & -- \\
+self-alignment & +0.0014 \\
+distillation & +0.0091 \\
+self-alignment + distillation & \textbf{+0.0108} \\
\bottomrule
\end{tabular}
\end{table}

Distillation provides the dominant gain, while self-alignment further improves results when
combined with distillation.

\subsubsection{Inference Optimizations}

\begin{table}[t]
  \caption{Ablation on Inference Optimizations ($<$500 ms latency)}
  \label{tab:ablation_inference}
  \small
  \centering
  \begin{tabular}{l l r}
    \toprule
    Configuration & Prefix Optimization & QPS (Items/s/GPU) \\
    \midrule
    \multirow{3}{*}{Raw-text}
        & None & 270 \\
        & Mult-Item Scoring & 275 \\
        & In-Batch Prefix Cache & 290 \\
    \midrule
    \multirow{3}{*}{Summarized, Pruned}
        & None & 1{,}650 \\
        & Multi-Item Scoring & 2{,}100 \\
        & In-Batch Prefix Cache & 2{,}200 \\
    \midrule
    \multirow{3}{*}{MixLM}
        & None & 3{,}000 \\
        & Multi-Item Scoring & 20{,}000 \\
        & In-Batch Prefix Cache & \textbf{22{,}000} \\
    \bottomrule
  \end{tabular}
\end{table}

MixLM combined with in-batch prefix caching yields the highest throughput, confirming the
importance of both input compression and shared-prefix amortization for high-QPS LLM ranking.

%% file: related_work.tex
\section{Related Works}

\subsection{LLMs for Semantic Search}

Extensive prior work studies the use of LLMs for semantic search. A widely adopted and
computationally efficient approach is the bi-encoder paradigm
\citep{relevance-biencoder,relevance-biencoder2}, which independently embeds queries and
documents into a shared vector space and is commonly deployed in production
\citep{facebook-ebr}. However, limited query--document interaction can miss fine-grained
semantics, motivating late-interaction models that introduce token-level matching
\citep{relevance-colbert}.

Cross-encoder models jointly encode queries and documents (or document lists), enabling richer
interaction modeling. LLM-based cross-encoders achieve strong ranking performance
\citep{sachan2022improving,zhuang2024setwise,zhang2024two,rankvicuna,relevance3}, but their
computational cost makes large-scale online serving challenging
\citep{related-pinterest,relevance-distill,relevance-distill2,rel-distill-reason}.

To address these limitations, prior work leverages LLMs indirectly. The \emph{LLM-as-judge}
paradigm uses LLMs to generate graded relevance labels and rationales to supervise efficient
student models
\citep{related-pinterest,relevance-tencent,relevance-distill3,relevance-distill4,rd-bench,rel-distill-reason},
while \emph{LLM-as-augmenter} approaches enrich training data via paraphrasing, summarization,
or hard-negative generation \citep{behdin2025scaling,relevance-neg}. Our work builds on these
directions while enabling direct deployment of LLM-based ranking through model--infrastructure
co-design.

\subsection{Mixed-Input LLMs for Recommendation Systems}

Mixed-input LLMs that combine text tokens with embedding inputs have been explored in
recommendation settings. Prior work replaces parts of text prompts with embeddings from
traditional models \citep{related2} or represents user interaction histories using embeddings
rather than text \citep{chen2024hllm}. These methods primarily target recommendation or offline
feature generation.

In contrast, we focus on relevance ranking, where nuanced semantic understanding is critical.
We present a complete training recipe for mixed-input ranking, including custom losses and
distillation from pure-text LLMs, and deploy the model directly for online inference with
reported A/B test results. We further describe serving-system optimizations required to scale
mixed-input ranking models in production.

Concurrent work by \citet{lin2025refrag} studies mixed-input LLMs for retrieval-augmented
generation, a setting distinct from ranking and evaluated primarily offline.

\subsection{Multimodal Large Language Model Training}

Multimodal LLMs commonly extend pretrained text models by injecting modality-specific embeddings
via lightweight adapters. Early systems such as Flamingo \citep{alayrac2022flamingo} and
Kosmos-1 \citep{huang2023language} treat non-text inputs as additional tokens to enable strong
vision--language performance.

Later work emphasizes minimal architectural changes by aligning pretrained encoders with LLM
embedding spaces. BLIP-2 \citep{li2023blip} uses a querying transformer to align vision features,
while LLaVA-style models \citep{liu2023visual} project CLIP features directly into the LLM token
space. MixLM follows this paradigm by injecting learned embedding tokens to replace long text
inputs, adapting multimodal techniques to high-throughput semantic ranking.